\documentclass[twocolumn,showpacs,preprintnumbers,superscriptaddress,amsmath,amssymb]{revtex4}

\usepackage{dcolumn}
\usepackage{bm}
\usepackage{amsmath}
\usepackage{multirow}
\usepackage{graphicx}

\begin{document}


 \newcommand{\re}{\mathop{\mathrm{Re}}}
 \newcommand{\im}{\mathop{\mathrm{Im}}}
 \newcommand{\D}{\mathop{\mathrm{d}}}
 \newcommand{\I}{\mathop{\mathrm{i}}}
 \newcommand{\E}{\mathop{\mathrm{e}}}
 \newcommand{\unite}[2]{\mbox{$#1\,{\rm #2}$}}
 \newcommand{\myvec}[1]{\mbox{$\overrightarrow{#1}$}}
 \newcommand{\mynor}[1]{\mbox{$\widehat{#1}$}}
 \newcommand{\rmsemit}{\mbox{$\tilde{\varepsilon}$}}
 \newcommand{\mean}[1]{\mbox{$\langle{#1}\rangle$}}

\title{Tunable subpicosecond electron bunch train generation \\
using a transverse-to-longitudinal phase space exchange technique}
\author{Y.-E Sun}
\affiliation{Accelerator Physics Center, Fermi National Accelerator
Laboratory, Batavia, IL 60510, USA}
\author{P. Piot}\affiliation{Accelerator Physics Center, Fermi National
Accelerator Laboratory, Batavia, IL 60510, USA} \affiliation{Northern Illinois Center for
Accelerator \& Detector Development and Department of Physics,
Northern Illinois University, DeKalb IL 60115,
USA}
\author{A. Johnson} \affiliation{Accelerator Division, Fermi National Accelerator
Laboratory, Batavia, IL 60510, USA}
\affiliation{Northern Illinois Center for Accelerator \& Detector Development and Department of Physics,
Northern Illinois University, DeKalb IL 60115, USA}
\author{A. H. Lumpkin} \affiliation{Accelerator Division, Fermi National Accelerator
Laboratory, Batavia, IL 60510, USA}
\author{T. J. Maxwell}\affiliation{Accelerator Physics Center, Fermi National
Accelerator Laboratory, Batavia, IL 60510, USA} \affiliation{Northern Illinois Center for
Accelerator \& Detector Development and Department of Physics,
Northern Illinois University, DeKalb IL 60115, USA}
\author{J. Ruan} \affiliation{Accelerator Division, Fermi National Accelerator
Laboratory, Batavia, IL 60510, USA}
\author{R. Thurman-Keup} \affiliation{Accelerator Division, Fermi National Accelerator
Laboratory, Batavia, IL 60510, USA}

\date{\today}

\begin{abstract}
We report on the experimental generation of a train of subpicosecond electron bunches. The bunch train  generation is accomplished using a beamline capable of exchanging the coordinates between the horizontal and longitudinal degrees of freedom.  An initial beam consisting of a set of horizontally-separated beamlets is converted into a train of bunches temporally separated with tunable bunch duration and separation.  The experiment reported in this Letter unambiguously demonstrates the conversion process and its versatility.
\end{abstract}
\maketitle

Recent applications of electron accelerators have spurred the demand for precise phase-space control schemes. In particular, electron bunches with a well-defined temporal distribution are often desired. An interesting class of temporal distribution consists of trains of bunches with subpicosecond duration and separation. Applications of such trains include the generation of super-radiant radiation~\cite{gover,bosco,ychuang} and the resonant excitation of wakefields in novel beam-driven acceleration methods~\cite{muggliprstab,jing}. To date there are very few methods capable of providing this class of beams reliably~\cite{muggli}. We have recently explored an alternative technique based on the use of a transverse-to-longitudinal phase space exchange method~\cite{piotAAC08,yineLINAC08}.  The method consists of shaping the beam's transverse density to produce the desired horizontal profile, the horizontal profile is then mapped onto the longitudinal profile by a beamline capable of exchanging the phase spaces between the horizontal and longitudinal degrees of freedom. Therefore the production of a train of bunches simply relies on generating a set of horizontally-separated beamlets upstream of the beamline, e.g., using a masking technique.
Considering an electron with coordinates ${\bf \widetilde{X}}\equiv (x,x'\equiv p_x/p_z, z, \delta\equiv p_z/\mean{p_z}-1)$ (here $p_x$, $p_z$ are respectively the horizontal and longitudinal momenta, $\mean{p_z}$ represents the average longitudinal momentum) 
in the four dimensional trace space, the $4\times 4$ transfer matrix $R$ associated to an ideal transverse-to-longitudinal phase-space-exchanging beamline is $2\times 2$-block anti-diagonal. Thus the beamline exchanges the emittances between the transverse and longitudinal degrees of freedom. The normalized horizontal root-mean-square (rms) emittance is defined as $\varepsilon_x^n \equiv \gamma \beta [\mean{x^2}\mean{x'^2} -\mean{x x'}^2]^{1/2}$,
where $\gamma$ is the Lorentz factor and $\beta \equiv \sqrt{1-\gamma^{-2}}$. A similar definition holds for the longitudinal degree of freedom. Phase-space-exchanging [or emittance-exchanging (EEX)] beamlines  were initially considered as a means to increase the luminosity in the B-factories~\cite{orlov}, mitigate instabilities in high-brightness electron beams~\cite{emma}, and improve the performance of single-pass free-electron lasers~\cite{emma2}.

A simple configuration capable of performing such a phase-space exchange consists of a horizontally-deflecting resonant cavity, operating in the TM$_{110}$ mode, flanked by two horizontally-dispersive sections henceforth referred to as ``doglegs"~\cite{kim}. Describing the beamline elements with their thin-lens-matrix approximation, an electron with initial trace space coordinates ${\bf X_0}$ will have its final coordinates ${\bf X}=R {\bf X_0}$. In particular the electron's final longitudinal coordinates $(z,\delta)$ are solely functions of its initial transverse coordinates $(x_0,x'_0)$~\cite{yine}
\begin{eqnarray}\label{eqn:phil}
\left\{ \begin {array}{ll} z =& -\frac{\xi}{\eta}x_0-\frac{L\xi-\eta^2}{\eta}x_0'\\
\delta =& -\frac{1}{\eta}x_0-\frac{L}{\eta}x_0'
\end{array},\right.
\end{eqnarray}
where $L$ is the distance between the dogleg's dipoles, and $\eta$ and $\xi$ are respectively the horizontal and longitudinal dispersions generated by one dogleg. Here the deflecting cavity is operated at the zero-crossing phase, i.e., the center of the bunch is not affected while the head and tail are horizontally deflected in opposite directions. The deflecting strength of the cavity $\kappa \equiv 2\pi|e| V_x/(\lambda c\mean{p_z})$ where $e$ is the electron charge, $\lambda$ is the wavelength of the TM$_{110}$ mode, and $V_x$ is the integrated maximum deflecting voltage, is chosen as $\kappa=-1/\eta$. The coupling described by Eq.~\ref{eqn:phil} can be used to arbitrarily shape the current or energy profile of an electron beam~\cite{piotPRSTAB}. \\

\begin{figure}[hhhhhhhhh!!!!!!!!!!!!!!]
\includegraphics[scale = 0.19]{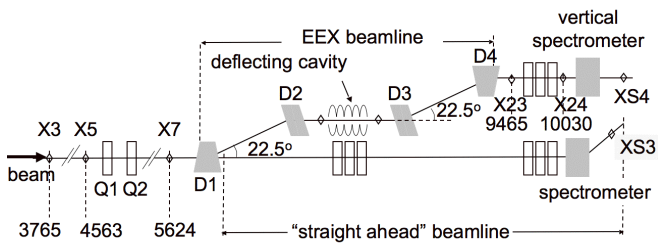}
\caption{Top view of the experimental setup displaying elements pertinent to the present experiment. The ``X" refers to diagnostic stations (beam viewers and/or multi-slit masks location), ``Q" the quadrupole magnets and ``D" the dipole magnets. Distances are in millimeters and referenced to the photocathode surface. The spectrometer dipole magnet downstream of the EEX beamline bends the beam in the vertical direction.\label{fig:beamline}}
\end{figure}

The experiment reported in this Letter uses the  $\sim 14$-MeV electron bunches produced by a radiofrequency (rf) photoemission electron source and accelerated in an rf superconducting  cavity at Fermilab's A0 Photoinjector~\cite{carneiro}. Downstream of the cavity, the beamline includes a set of quadrupole and steering dipole magnets, and beam diagnostics stations before splitting into two beamlines as shown in Fig.~\ref{fig:beamline}.

The ``straight ahead" beamline incorporates a horizontally-bending spectrometer equipped with a Cerium-doped Yttrium Aluminum Garnet (Ce:YAG) scintillating screen (labeled as XS3 in Fig.~\ref{fig:beamline}) to measure the beam's energy distribution. The horizontal dispersion at the XS3 location is $317$~mm.
\begin{table}[h!]
\caption{\label{tab:operating} Typical initial beam parameters measured before emittance exchange. The Courant-Snyder (C-S) parameters are $\alpha_x\equiv -\mean{x x'}/\varepsilon_x$ and $\beta_x\equiv \mean{x^2}/\varepsilon_x$, where $\varepsilon_{x}\equiv  \varepsilon_{x}^n /(\beta\gamma) $ is the geometric emittance. }
\begin{center}
\begin{tabular}{l c c c}
\hline \hline Parameter                   & Symbol     &      Value       & Units  \\ \hline
energy                        & $E$           &  14.3 $\pm$ 0.1     & MeV \\
charge                       & $Q$    & $550\pm 30$    & pC     \\
rms duration              & $\sigma_t$ &        4.0 $\pm$ 0.3 & ps     \\
horizontal emit.         & $\varepsilon_x^n$         &  $4.7\pm 0.3$   & $\mu$m  \\
rms frac. energy spread  &  $\sigma_{\delta}$               &  0.06  $\pm$ 0.01  & \%     \\
horizontal C-S param. & $(\alpha_{x},\beta_x)$ & ($1.2\pm 0.3$,$14.3\pm 1.6$) & (--,m)    \\
\hline \hline
\end{tabular}
\end{center}
\end{table}

The other beamline, referred to as the EEX beamline, implements the double-dogleg setup described above~\cite{koeth0} and has been used to explore emittance exchange~\cite{amber}. The doglegs consist of dipole magnets with $\pm 22.5^{\circ}$ bending angles and each generates horizontal and longitudinal dispersion of $\eta\simeq -33$~cm and $\xi\simeq -12$~cm, respectively~\cite{footnote}.  The deflecting cavity is  a liquid-Nitrogen--cooled five-cell copper cavity operating on the TM$_{110}$ $\pi$-mode at 3.9~GHz~\cite{koeth}. The section downstream of the EEX beamline includes three quadrupoles,  beam diagnostics stations and a vertical spectrometer. The dispersion generated by the spectrometer at the XS4 Ce:YAG screen is  $944 $~mm. The temporal distribution of the electron bunch is diagnosed via the coherent transition radiation (CTR) transmitted through a single-crystal quartz window as the beam impinges an aluminum foil at X24. The CTR is sent through a Michelson autocorrelator~\cite{tr} and the autocorrelation function is measured by a liquid helium-cooled bolometer which is used as the detector of the autocorrelator. The CTR spectrum is representative of the bunch  temporal distribution provided $\sigma_{\perp} \ll \gamma \sigma_z$ where $\sigma_z$ and $\sigma_{\perp}$ are respectively the rms bunch length and transverse size at the CTR radiator location (the beam is assumed to be cylindrically symmetric at this location).  In the present experiment the beam was focused to an rms spot size of $\sigma_{\perp} \simeq 400$~$\mu$m at X24. Imperfections due to the frequency-dependent transmissions of the THz beamline components alter the spectrum of the detected CTR and limit the resolution to $\sim 200$~fs.

\begin{figure}[hhhhh!!!!]
\includegraphics[scale = 0.62]{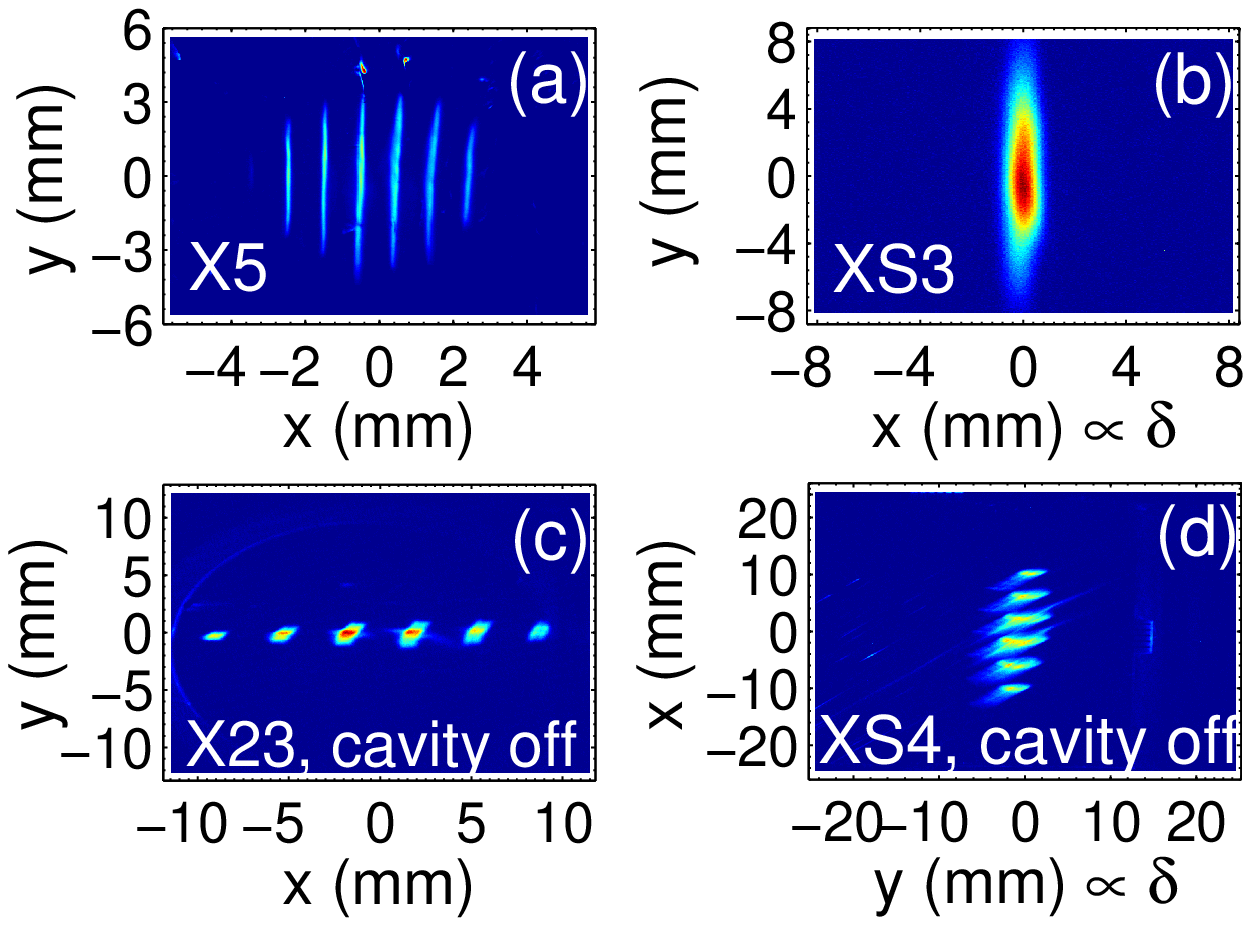}
\includegraphics[scale = 0.62]{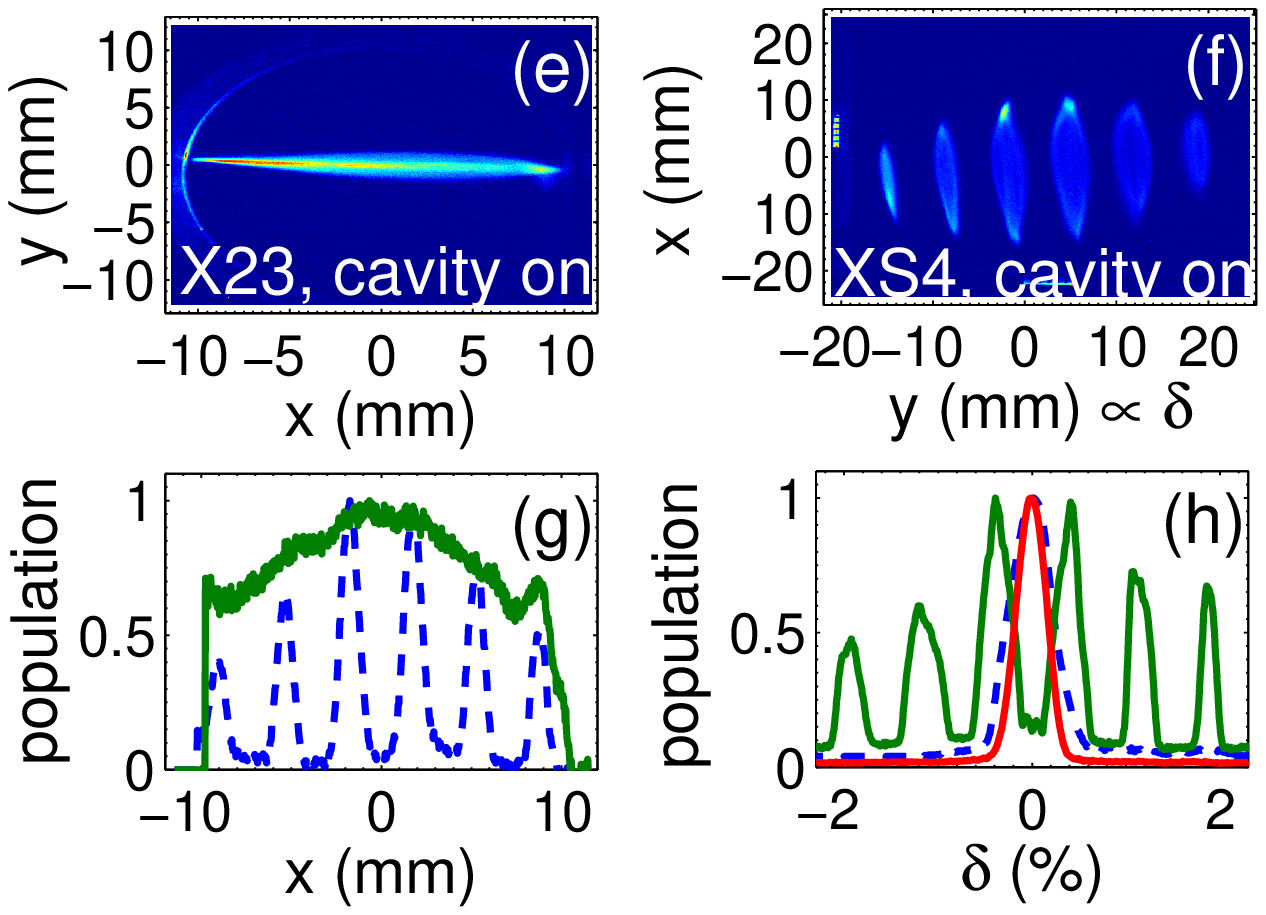}
\caption{Transverse initial beam density at X5 (a), XS3 (b) and corresponding final beam density at X23 with deflecting cavity off (c) and on (e), and at XS4 with deflecting cavity off (d) and on (f). The corresponding relevant intensity-normalized horizontal profile at X23 (g) and fractional energy spread (h) profiles  obtained from XS3 (red) and XS4 for the cases when the cavity is on (green) and off (dashed blue) are also displayed. \label{fig:dppanalysis}}
\end{figure}

For the proof-of-principle experiment reported here, a number of horizontally-separated beamlets were  generated by passing the beam through a set of vertical slits at X3. The measured parameters for the incoming beam are gathered in Table~\ref{tab:operating}. The multislit mask, nominally designed for single-shot transverse emittance measurements, consists of 48~$\mu$m-wide slits made out of a 3-mm-thick tungsten plate. The slits are separated by 1~mm. Less than 5~\% of the incoming beam is transmitted through the mask. Up to 50 electron bunches repeated at 1~MHz were used to increase the signal-to-noise ratio of the measurements.

The beam was first diagnosed in the straight-ahead line to ensure that horizontal modulations are clearly present and there are no energy modulations [Fig.~\ref{fig:dppanalysis} (a) and (b)]. It was then transported through the EEX beamline with the deflecting cavity turned off. The transverse modulation was still observable at X23 but no energy modulation could be seen at XS4 as shown in Fig.~\ref{fig:dppanalysis} (c), (d), (g) and (h). Powering the cavity to its nominal deflecting voltage ($V_{x}\simeq 720$~kV) resulted in the suppression of the transverse modulation at X23 and the appearance of an energy modulation at XS4 [Fig.~\ref{fig:dppanalysis} (e), (f), (g) and (h)]. These observations clearly demonstrate the ability of the EEX beamline to convert an incoming transverse density modulation into an energy modulation. In the present measurement the incoming horizontal Courant-Snyder (C-S) parameters at the EEX beamline entrance were empirically tuned for energy and time modulation in the beam by setting the current of quadrupole magnets $Q_1$ and $Q_2$ to respectively 1.6~A and -0.6~A.

 \begin{figure}[hhhhhh]
\includegraphics[scale = 0.55]{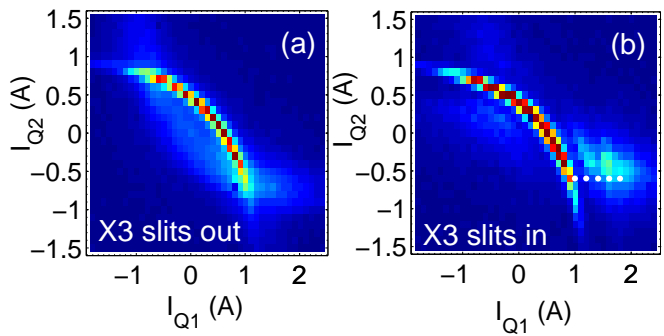}
\caption{Total normalized CTR energy detected at X24 as a function of quadrupole magnets currents $I_{Q_1}$ and $I_{Q_2}$ with X3 slits out(a) and in (b) the beamline. The bolometer signal is representative of the inverse of the bunch duration $\sigma_t$. The white dots in (b) indicate loci where more detailed measurements were performed; see Fig.~\ref{fig:deltaT}. \label{fig:totalCTR}}
\end{figure}

\begin{figure}[hhhh!!!!]
\includegraphics[scale = 0.6]{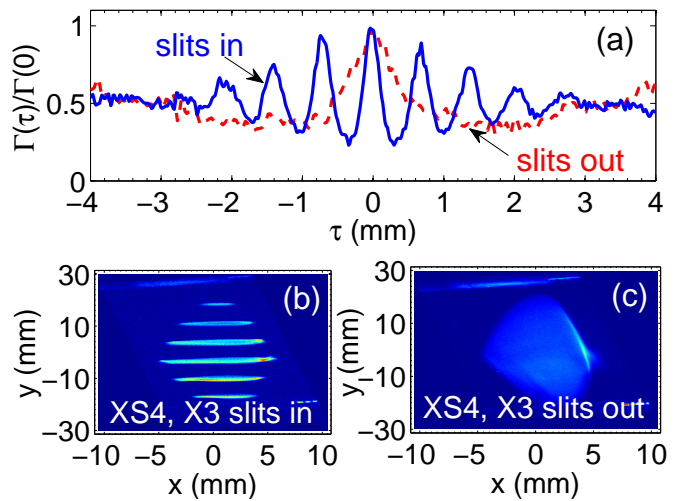}
\caption{ (a) Normalized autocorrelation function $\Gamma(\tau)/\Gamma(0)$ of the CTR signal (a) recorded with (solid) and without (dashed) the X3 slits inserted as a function of the optical path difference $\tau$. The corresponding beam transverse densities at XS4 appear in (b) and (c).  The vertical axis on the bottom image is proportional to the beam's fractional momentum spread ($\delta$). The nominal bunch charge is $550\pm 30$~pC and reduces to $\sim 15 \pm 3$~pC when the slits are inserted. \label{fig:autocorr}}
\end{figure}

To characterize the expected temporal modulations we detect and analyze the CTR  emitted as the beam impinges the X24 aluminum foil~\cite{saxon}. The total CTR energy detected within the detector bandwidth $[\omega_l,\omega_u]$ and angular acceptance  increases as the bunch duration $\sigma_t \equiv \sigma_z/c$ decreases. In the limit $\omega_l \ll \sigma_t^{-1} \ll \omega_u$, the total radiated energy is inversely proportional to the rms bunch duration~\cite{piotvelo}. The final longitudinal C-S parameters downstream of the EEX beamline can be varied by altering the initial horizontal C-S parameters using the quadrupole magnets $Q_1$ and $Q_2$.  Figure~\ref{fig:totalCTR} shows the detected CTR energy as a function of quadrupole magnet currents for the cases without (a) and with (b) intercepting the beam with the X3 multislit mask.  The two plots illustrate the ability to control the final bunch length (as monitored by the CTR power detected at X24) using the EEX technique. The insertion of the multislit mask results in the appearance of a small island of coherent radiation at the lower right corner of Fig.~\ref{fig:totalCTR} (b). The corresponding autocorrelation functions $\Gamma(\tau)$ (where $\tau$ is the optical path difference) recorded by the bolometer for the quadrupole magnets currents $(I_{Q1},I_{Q2})=$(1.6~A,-0.6~A) are shown in Fig.~\ref{fig:autocorr} (a) with and without inserting the multislit mask. When the multislit mask is inserted the autocorrelation function is multipeaked indicating a train of bunches is produced. For this particular case a train of $N=6$ bunches with unequal peak intensity are produced resulting in an autocorrelation function with $2N-1=11$ peaks. The measured separation between the bunches is  $\Delta z =762 \pm 44 $~$\mu$m.  It should be noted that the two autocorrelations shown in Fig.~\ref{fig:autocorr} correspond to very different charges and longitudinal space charge effects influence the bunch dynamics and result in different final longitudinal C-S parameters. In addition, the low frequency limit of the CTR detection system prevents the accurate measurement of autocorrelation functions of bunches with rms length larger than $\sim 500$$\mu$m~\cite{TimM}.

\begin{figure}[t]
\includegraphics[scale = 0.6]{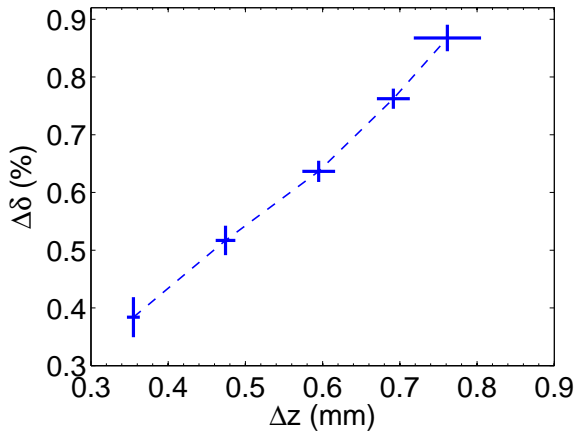}
\caption{Fractional momentum spread separation $\Delta \delta$ versus time separation $\Delta t$ between the bunches within the train for different initial beam conditions. The different data points are obtained from the autocorrelation functions recorded for settings   $I_{Q1}=1.0$, 1.2, 1.4, 1.6, and 1.8~A (from left to right) shown as white dots in Fig.~\ref{fig:totalCTR}. The current  $I_{Q2}$ is kept constant at $-0.6$~A. \label{fig:deltaT}}
\end{figure}

In addition, varying the settings of the quadrupole magnets provide control over the final longitudinal phase space time-energy correlation. The correlation can be measured as the ratio of the peak separation along the longitudinal coordinate and  the energy ${\cal C} = \mean{z\delta}/\mean{z^2} \simeq \Delta \delta / \Delta t$. These measurements are presented in Fig.~\ref{fig:deltaT} for different quadrupole magnet settings. As shown in Fig.~\ref{fig:deltaT}, the technique can provide a tunable bunch spacing ranging from $\sim 350$ to $760$~$\mu$m given an initial slit spacing of 1~mm by adjusting one quadrupole magnet strength (Q1) only. For $\Delta z \simeq 350$~$\mu$m (corresponding to $\Delta t=\Delta z/c \simeq 1.2$~ps), the autocorrelation has a 100\% modulation implying that the bunches within the train are fully separated. Assuming the bunches follow a Gaussian distribution, their estimated rms duration is $< 300$~fs (this estimate includes the finite resolution of our diagnostics). Variation of both Q1 and Q2 quadrupole magnet strengths can generate even shorter bunch separations, however our current measurement system has limited sensitivity in the shorter wavelength region, resulting in less than 100\% modulation in the autocorrelation curve.

%

In summary we have experimentally demonstrated that an incoming phase space modulation in the horizontal coordinate can be converted into the longitudinal phase space using an EEX beamline. The method was shown to produce  energy- and time-modulated bunches arranged as a train of subpicosecond bunches with variable spacing. This proof-of-principle experiment also provides an unambiguous demonstration of the main property of the EEX beamline to exchange the phase space coordinates between the horizontal and longitudinal degrees of freedom. The technique experimentally demonstrated in this Letter can be used to tailor the current and energy profile of electron beams and could have applications in novel beam-driven acceleration techniques, compact short-wavelength accelerator-based light sources, and ultra-fast electron diffraction.

We are indebted to E. Harms, E. Lopez, R. Montiel, W. Muranyi, J. Santucci, C. Tan and B. Tennis for their technical supports. We thank M. Church, H. Edwards, and V. Shiltsev for their interest and encouragement. The work was supported by the Fermi Research Alliance, LLC under the U.S. Department of Energy Contract No. DE-AC02-07CH11359, and by Northern Illinois University under the US Department of Energy Contract No. DE-FG02-08ER41532.


\begin{thebibliography}{9}
\bibitem{footnote}In our convention, the head of the bunch is for $z<0$.
\end{thebibliography}
\end{document}